\documentclass[journal]{IEEEtran}

\usepackage{amssymb}
\usepackage{amsmath}
\usepackage{makeidx}
\usepackage{multicol}
\usepackage[abs]{overpic}
\usepackage{tabularx}
\usepackage[numbers,sort&compress]{natbib}
\usepackage{amsfonts}
\usepackage{color}
\usepackage{array,color}
\usepackage{tabularx}
\usepackage{multirow}
\usepackage{graphics}
\usepackage{epsfig}
\usepackage{graphicx}
\usepackage{subfigure}
\usepackage{makeidx}
\usepackage{multicol}
\usepackage{amsthm}
\usepackage{bm}
\usepackage{booktabs}
\usepackage{ulem}
\usepackage{enumerate}
\usepackage{subfig}
 \usepackage{float}
\usepackage{epstopdf}

\newtheorem{proposition}[theorem]{Proposition}

\begin{document}
\title{	A Data-driven Approach to Multi-event Analytics in Large-scale Power Systems Using Factor Model}
\author{Fan Yang, Xing He, Robert Caiming Qiu,~\IEEEmembership{Fellow,~IEEE} and Zenan Ling}
\maketitle
\begin{abstract}
Multi-event detection and recognition in real time is of challenge for a modern grid as its  feature is usually non-identifiable.  Based on factor model, this paper porposes a data-driven method as an alternative solution under the framework of random matrix theory. This method maps the raw data into a high-dimensional space with two parts: 1) the principal components (factors, mapping event signals); and 2) time series residuals (bulk, mapping white/non-Gaussian noises). The spatial information is extracted form factors, and the termporal infromation from residuals. Taking both  spatial-tempral correlation into account, this method is able to reveal the multi-event: its components and their respective details, e.g., occurring time. Case studies based on the standard IEEE 118-bus system validate the proposed method.
\end{abstract}

\IEEEpeerreviewmaketitle
\section{Introduction}
\IEEEPARstart{F}{or} a large-scale power system, multiple events can hardly be identified properly as it is difficult to distinguish the features of multi-event from the ones of single-event. The multi-event poses a more serious threat to the systems: it can hardly be identified, and thus be addressed, which may lead to a wide-spread blackout.

This paper proposes a statistical, data-driven solution, rather than its deterministic, empirical or model-based counterpart, to solve the problem given above.  The study is built upon our previous work in the last several years. See Section~\ref{sec:RelatedW} for details.

\subsection{Contribution}
This paper, based on random matrix theory (RMT), proposes a high statistical tool, namely, factor model, for multi-event detection and recognition in a modern grid. This paper extracts the spatial and temporal information from the massive raw data, respectively, in the form of principal components (factors) and residuals (bulk). The factors map event signals, and the residuals map white/non-Gaussian noises.

The proposed method can be used for multi-event analytics effectively. To the factors, we experimentally obtain that there is a linear relationship between the number of factors and the event number of the multi-event.
To the residuals, on the other hand, we extract their information rather than simply assuming it to be identically independent “pure white noise” as \cite{Xu2015A}. Time series information contained in noise, together with the spatial information in factors, reveals the multi-event status. The proposed method is practical for real-time analysis.

Besides, the proposed solution is model-free, requiring no knowledge of topologies or parameters of the power system
\cite{Miao2014The,Zhang20135Ws,Mayilvaganan2013A}, and able to handle non-Gaussian noises. To the best of our knowledge, it is the first time to propose an algorithm  aiming at multi-event detection based on random matrix theory in the field of power systems.

\subsection{Related Work}
\label{sec:RelatedW}

 In our previous work, a universal architecture with big data analytics is proposed \cite{He2017A} and is applied for anomaly detection \cite{He2015An,Chu2016Massive}. Little work, howerve, has been done to multi-event analytics in a complex situation.
Current researches on event analysis are mainly model-based, aiming at single-event analytics. They may not be suitable for real-time analytics in a complex situation \cite{Smith2009Event}. Some other methods adopt graph theory \cite{Soltan2014Cascading,He2011A}; the methods strongly depend on the structure of the power system.

Some data-driven methods for event analysis are proposed recently \cite{Xie2014Dimensionality} and applied for multi-event analytics \cite{Rafferty2016Real}.

Rafferty utilizes principal component analysis (PCA) for real-time multi-event detection and classification in \cite{Rafferty2016Real}. In his approach, a threshold of cumulative percentage of total variation is selected in advance. Then, the number of principal components is determined according to the threshold mentioned above. The threshold is selected empirically and subjectively. Besides, for other supervised tools, like deep learning \cite{Wang2016Deep} and kernel-based algorithms \cite{Naderian2016An}, which are hot-spot to data-driven approaches, the same problem is inevitable. The deep learning algorithms automatically select the features from the massive datasets. This is one big advantage of deep learning over our paradigm. Our paradigm, however, has the advantage of transparency in that our results are provably. Also, our paradigm is deeply rooted in random matrix theory.

Nowadays, high-dimensional factor model has been actively studied and already successfully applied in statistics \cite{Forni2016Dynamic} , econometrics \cite{Yeo2016Random} and biology \cite{Sun2016A}.

\section{Theory Foundation and Data Processing}
The  frequently used notations are given in Table 1.

\subsection{Random Matrix Theory and Spectral Analytics}

Random matrices have been an important issue in multi-variate statistical analysis since the landmark work of Wigner and Wishart \cite{Wigner1951On}, motivated by problems in quantum physics. Factor model, on the other side, can be used to identify non-random properties (in the form of spikes/outliers) which are deviations from the universal predictions (in the form of bulk) \cite{Plerou2012Universal,LAURENT2000RANDOM}. To be specific, the eigenvalues of covariance matrix (spectrum) can usually be divided into two parts: a bulk and several spikes. The bulk represents the noises, while the spikes represent the signals, namely, factors.

In previous work, noises are usually assumed to be identically independent in power systems, namely, white noises. However, "no information" or "pure noise" assumption  is invalid in practice. For instance, for a certain PMU, there exits time correlation  between the measured voltage magnitude data of adjacent sampling points \cite{Chevalier2016Identifying,Xu2005The}. The time correlation is non-ignorable, especially for a large inter-connected system! This paper formulars the noises using time series analytics.

\begin{table}[ht]
	\begin{center}
		\caption{{\label{table1}} Some Frequently Used Notations}
		\begin{tabular}{p{1.5cm}|p{6.5cm}}
			\toprule
			\textbf{Notations} & \textbf{Means}  \\
			\midrule

			$\bm{X},\bm{x},{x_{i,j}}$   &  a matrix, a vector, an entry of a matrix  \\
			$\mu (x),\sigma (x)$ &  mean, variance for $x$  \\
			$\Omega$   & raw data source  \\
			${\mathbb{C}^{N \times T}}$   &  $N \times T$ dimensional complex space \\
			$N,T$   &  the row  and column size of moving split-window  \\
			$n$   &  the number of measurable status variables  \\
			${t_i}$   &  the sampling time  \\
			${\hat {\bm{X}}}$   &  a raw data matrix  \\
			${\tilde {\bm{X}}}$   & a standard non-Hermitian matrix  \\
			$p$   &  the number of factors\\
			$b$   &  the covariance structure of residualss\\
			$z$   &  a complex eigenvalue, $z=\lambda+i\varepsilon$\\
			$\hat p,\hat \theta$   &  the estimated value of $p$, $\theta$ \\
			\bottomrule
		\end{tabular}
	\end{center}
\end{table}

Reference \cite{Zhang2007Spectral} provides a fundamental theory to estimate noises, and formulates them as:

\begin{equation}
\label{eq1}
\bm{U} = {\bm{A}_N}^{1/2}\bm{G}{\bm{B}_T}^{1/2}
\end{equation}

where $\bm{G}$ is an $N \times T$ matrix with i.i.d (identically independent distribution) Gaussian entries, and $\bm{A}_N$ and $\bm{B}_T$ are $N \times N$ and $T \times T$ symmetric non-negative definite matrices, representing cross- and auto- covariances, respectively. For more details about the model, please refer to Appendix A.
On the other side, the spikes (deviating eigenvalues) of the spectrum map the event signals. They represent dominant information for system operating status.
\subsection{Data Processing}
Massive raw data can be represented by matrix naturally \cite{Qiu2013Cognitive}. In a power system, assume that there are $n$ kinds of measurable variables. At sampling time $t_0$, we arrange the measured data of these variables in the form of a column vector $\hat {\bm{x}}({t_0}) = {({{\hat x}_{t_0,1}},{{\hat x}_{t_0,2}}, \cdots ,{{\hat x}_{t_0,n}})^H}$ \cite{Donoho2000High}. Then, arranging the column vectors $\hat x({t_i})$ in chronological order ($i=1,2,\cdots$), we obtain
raw data source $\Omega$.

For the raw data source $\Omega$, we can cut off any arbitrary part, e.g., size of $N \times T$, at any time, e.g., samping time $t_i$ , forming  ${\hat {\bm{X}}_{ti}} \in {\mathbb{C}^{N \times T}}$ as

\begin{equation}
\label{eq2}
\hat {\bm{X}}({t_i}) = (\hat {\bm{x}}({t_{i - T + 1}}),\hat {\bm{x}}({t_{i - T + 2}}), \cdots ,\hat {\bm{x}}({t_i}))
\end{equation}
where $\hat {\bm{x}}({t_j}) = ({\hat x_{t_j,1}},{\hat x_{t_j,2}}, \cdots ,{\hat x_{{t_j,N}}})^H$  is measured data at sampling time ${t_j}$ ($j=1,2,\cdots,T$).
It is worth noting that $T$ is the length of the moving split-window. If we keep the  last sampling time as the current time, with the moving split-window, the real-time analytics is conducted.

Then, we convert the raw data matrix ${\hat {\bm{X}}_{ti}}$ obtained at each sampling time $t_i$ into a standard non-Hermitian matrix ${\tilde {\bm{X}}_{ti}}$ with the following algorithm.
\begin{equation}
\label{eq3}
{\tilde x_{i,j}} = ({\hat x_{i,j}} - \mu ({\hat {\bm{x}}_i})) \times \frac{{\sigma ({{\tilde {\bm{x}}}_i})}}{{\sigma ({{\hat {\bm{x}}}_i})}} + \mu ({\tilde {\bm{x}}_i})
\end{equation}
where ${\hat {\bm{x}}_i} = ({\hat x_{i,1}},{\hat x_{i,2}}, \cdots ,{\hat x_{i,T}})$, $\mu ({\tilde {\bm{x}}_i}) = 0$, $\sigma ({\tilde {\bm{x}}_i}) = 1$,  $i = 1,2, \cdots ,N$ and $j = 1,2, \cdots ,T$.

In the following section, ${\tilde {\bm{X}}_{ti}}$ is used to analyze the factors and noises at sampling time $t_i$.

\section{Factor Model Algorithm}
For a certain window, e.g. the one obtained at sampling time $t_i$, we aim to decompose the  standard non-Hermitian matrix ${\tilde {\bm{X}}_{ti}}$, as given in \eqref{eq3}, into factors and residuals as follows:
\begin{equation}
\label{eq4}
{\tilde {\bm{X}}_{ti}} =  {{{\bm{L}}_{ti,j}}{{\bm{F}}_{ti,j}} + {{\bm{U}}_{ti}}}
\end{equation}

where $p$ is the number of factors, ${{{\bm{F}}_{ti,j}}}$ is the $j-$th  factor, ${{{\bm{L}}_{ti,j}}}$ is the corresponding loading, ${{U_{ti}}}$ is the residual. Usually, only ${\tilde {\bm{X}}_{ti}}$ is available, while ${{{\bm{L}}_{ti,j}}}$, ${{{\bm{F}}_{ti,j}}}$ and ${{{\bm{U}}_{ti}}}$ need to be estimated.

Factor model aims  to simultaneously provide estimators of factor number and correlation structures in residuals. We turn the parameter-estimation problem into a minimum-distance problem. Specifically, we consider a minimum distance between the experimental spectral distribuion $\rho _{\text{real}}(p)$ and the theoretical spectral distribuion $\rho _{\text{model}}(b)$. The experimental one $\rho _{\text{real}}(p)$, depending on the sampling data, is obtained as empirical eigenvalue density (EED) of $C_{\text{real}}^{(p)}$ in \eqref{eq7}, and the theoretical one $\rho _{\text{model}}(b)$, based on Sokhotskys formlua, is given as $\rho _{\text{model}}(\lambda ;b)$ in \eqref{eq8}.
As a result, we turn the factor model estimation into a classical optimziation as \begin{equation}
\label{eq5}
\{ \hat p,\hat \theta \}  = \arg \min D(\rho _{\text{real}}(p),{\rho _{\text{model}}}(\theta ))
\end{equation}

where $D$ is a spectral distance measure or loss function. The solution of this minimization problem gives the number of factors, in the form of $\hat p$, and the parameters for the correlation structure of the residuals, in the form of $\hat \theta$.

\subsection{Principal Component Estimation :$\rho _{\mathrm{real}}(p)$}
The first step is to generate $p$-level empirical residuals, by substracting $p$ largest principal components according to \eqref{eq4}.
\begin{equation}
\label{eq6}
{\hat {\bm{U}}^{(p)}} = {\tilde {\bm{X}}_{ti}} - {\hat {\bm{L}}^{(p)}}{\hat {\bm{F}}^{(p)}}
\end{equation}

where ${\hat {\bm{F}}^{(p)}}$ is a $p \times T$ matrix of $p$ factors, each row of which is a $j$-th $(j = 1, \cdots ,p)$ principal component from ${{\tilde {\bm{X}}}_{ti}}^T{{\tilde {\bm{X}}}_{ti}}$, $\hat {\bm{L}}^{(p)}$  is an $N \times p$ matrix of factor loadings, estimated by multivariate least squares regression of $\tilde {\bm{X}}_{ti}$ on ${\hat {\bm{F}}^{(p)}}$.

Then the covariance matrix from $p$-level residuals is obtained as
\begin{equation}
\label{eq7}
\bm{C}_{\text{real}}^{(p)} = \frac{1}{T}{{\hat {\bm{U}}}^{(p)}}{{\hat {\bm{U}}}^{{{(p)}^T}}}
\end{equation}
The subscript ``real" indicates that $\bm{C}_{\text{real}}^{(p)}$ is obtained from real data. The steps can be summarized as follows:
\begin{table}[H]
	\begin{center}
		\begin{tabular}{p{8.5cm}}
			\toprule
			\textbf{Steps of Calculating $\rho _{\mathrm{real}}(p)$}\\
			\midrule
			1.Calculate ${{\hat {\bm{F}}  }^{(p)}}$: each row of which is a $j$-th principal component from correlation matrix of ${\tilde {\bm{X}}_{ti}}$, i.e. ${{\tilde {\bm{X}}}_{ti}}^T{{\tilde {\bm{X}}}_{ti}}$; denote as: ${{\hat {\bm{F}}}^{(p)}} = {({f_1},{f_2}, \cdots ,{f_p})^T}$.\\
			2.Conduct least squares regression of ${\tilde {\bm{U}}}_{ti}$ on ${\hat {\bm{F}}^{(p)}}$: ${{\hat {\bm{L}}}^{(p)}} = {{\tilde {\bm{X}}}_{ti}}{{\hat {\bm{F}}}^{{{(p)}^T}}}$.\\
			3.Calculate $p$-level residual: ${{\hat {\bm{U}}}^{(p)}} = {{\tilde {\bm{X}}}_{ti}} - {{\hat {\bm{L}}}^{(p)}}{{\hat {\bm{F}}}^{(p)}}$.\\
			4.Calculate covariance matrix from $p$-level residual: \\${\bm{C}}_{\text{real}}^{(p)} = \frac{1}{T}{{\hat {\bm{U}}}^{(p)}}{{\hat {\bm{U}}}^{{{(p)}^T}}}$.\\
			5.Calculate the empirical eigenvalue density of ${\bm{C}}_{\text{real}}^{(p)}$.\\
			\bottomrule
		\end{tabular}
	\end{center}
\end{table}
\subsection{Modeling Covariance of Residuals: $\rho _{\mathrm{model}}(\theta )$}
In section ${\rm I}{\rm I}$, we consider residuals as time series, which is represented by \eqref{eq1}. The $\rho _{\text{model}}(\theta )$, however, is difficult to be obtained, since the limiting distribution of general ${A_N}$ and ${B_T}$ cost too much calculation resource via Stieltjes transform in \cite{Zhang2007Spectral} Fortunately, a recent work by \cite{Zdzis2010A} provides an analytic derivation of limiting spectral density using free random variable techniques. This paper uses the results of \cite{Zdzis2010A} to calculate $\rho _{\text{model}}( \bullet )$. If we assume that the cross-correlations \cite{Zhang2007Spectral}, i.e. ${\bm{A}_N}$, are effectively removed by the factors, then,
the cross-correlations among the normalized residuals are negligible: ${\bm{A}_N} \approx {\bm{I}_{N \times N}}$. Under this assumption, only the auto-correlations, i.e. ${\bm{B}_T}$, left. The ${\bm{B}_T}$ is  in the forms of exponential decays with respect to time lags, as: ${({\bm{B}_T})_{i,j}} = {b^{\left| {i - j} \right|}}$. As a result, the $\rho _{\text{model}}({\theta _{{A_N}}},{\theta _{{B_T}}})$ is replaced by $\rho _{\text{model}}(b)$.

This enables us to calculate the modeled spectral density, $\rho _{\text{model}}(b)$, much more easily. It can be done through the free random variable techniques proposed in \cite{Zdzis2010A} (Refer to Appendix A for analytic derivation). The steps can be summarized as follows:
\begin{table}[H]
	\begin{center}
		\begin{tabular}{p{8.5cm}}
			\toprule
			\textbf{Steps of Calculating $\rho _{\mathrm{model}}(b)$}\\
			\midrule
			1.Get the mean spectral density from Green Function G(z) by using Sokhotsky’s formula:
			\begin{equation}
			\label{eq8}
			\rho _{\text{model}}(\lambda ;b)
			=  - \frac{1}{\pi }\mathop {\lim }\limits_{\varepsilon  \to {0^ + }} {\mathop{\rm Im}\nolimits} {G_c}(\lambda  + i\varepsilon )
			\end{equation}\\
			2. Green Function $G(z)$ can be obtained from the Moments Generating Function $M(z)$
			\begin{equation}
			\label{eq9}
			M(z) = zG(z) - 1
			\end{equation}\\
			3. Solve the polynomial equation for $M = M(z)$ $(a = \sqrt {1 - {b^2}} )$ and $c = N/T$ (a 6th-order polynomial equations for $\rho _{\text{model}}({\theta _{{A_N}}},{\theta _{{B_T}}})$):
			\begin{center}
				${a^4}{c^2}{M^4} + 2{a^2}c( - (1 + {b^2})z + {a^2}c){M^3} + $
			\end{center}
			\begin{equation}
			\label{eq10}
			({(1 - {b^2})^2}{z^2} - 2{a^2}c(1 + {b^2})z + ({c^2} - 1){a^4}){M^2} - 2{a^4}M - {a^4} = 0
			\end{equation}\\
			\bottomrule
		\end{tabular}
	\end{center}
\end{table}

With the above procedure, we can rewrite \eqref{eq5} as
\begin{equation}
\label{eq05}
\{ \hat p,\hat b \}  = \arg \min D(\rho _{\text{real}}(p),{\rho _{\text{model}}}(b ))
\end{equation}

\subsection{Distance Measure}
Since the empirical spectrum contains spikes, a distance measure which is sensitive to the presence of spikes should be given.
This paper uses Jensen-Shannon divergence, which is a symmetrized version of Kullback-Leibler divergence.
\begin{equation}
\label{eq11}
{D_{\text{JS}}}(P\left\| Q \right.) = \frac{1}{2}{D_{\text{KL}}}(P\left\| M \right.) + \frac{1}{2}{D_{\text{KL}}}(Q\left\| M \right.)
\end{equation}

where $P$ and $Q$ are probability densities, $M = \frac{1}{2}(P + Q)$, and ${D_{\text{KL}}}(P\left\| Q \right.)$ is the Kullback-Leibler divergence defined by ${D_{\text{KL}}}(P\left\| Q \right.) = \sum\limits_i {{P_i}\log \frac{{{P_i}}}{{{Q_i}}}}$ . Note that the Kullback-Leibler distance becomes larger if one density has a spike at a point while the other is almost zero at the same point. Refer to Appendix B for more details.

\begin{figure}[!htp]
	{
		\includegraphics[width=0.55\textwidth]{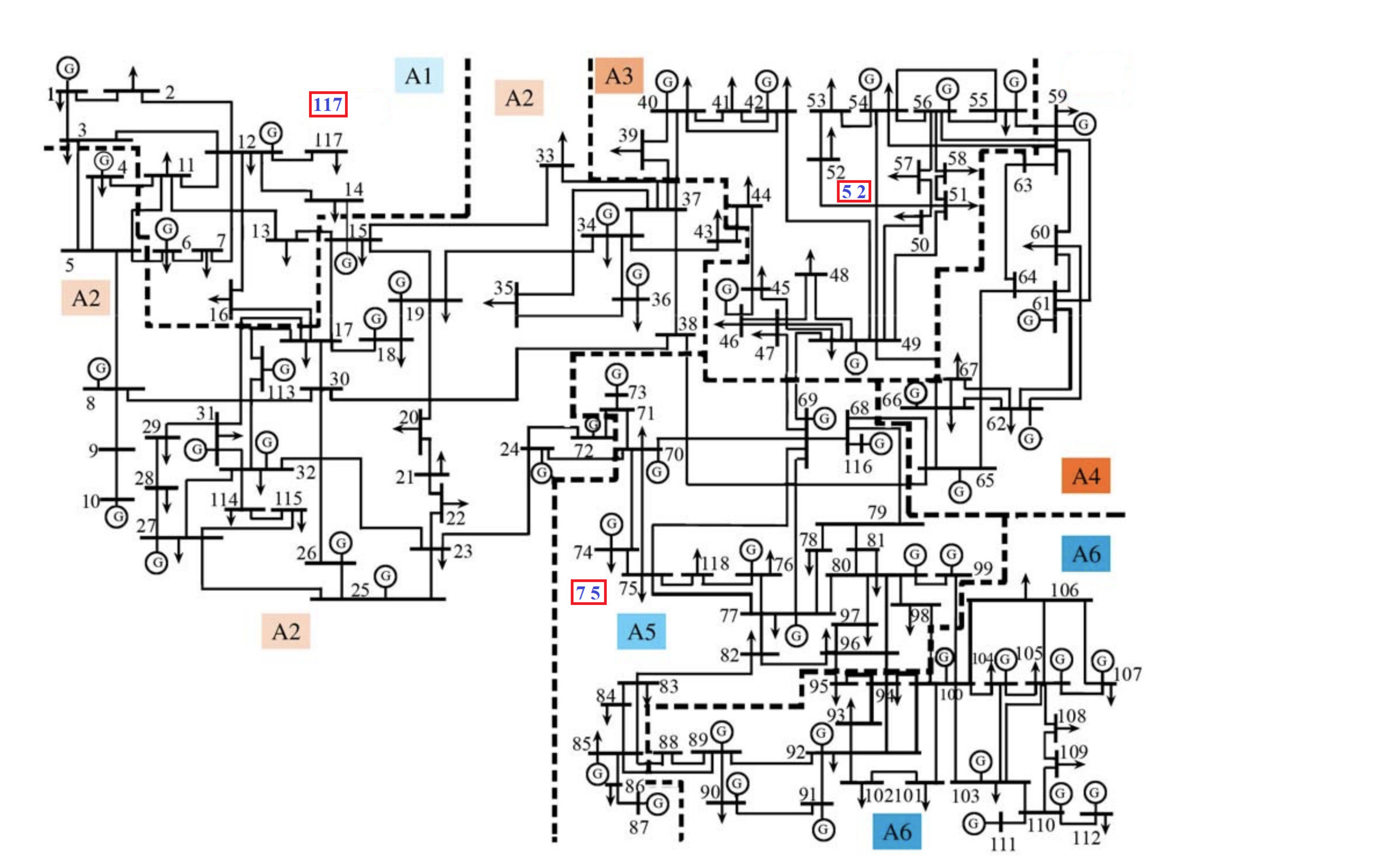}
	}
	\caption{{\label{fig1}} Topology of the Standard IEEE 118-bus System.}
\end{figure}

\section{CASE STUDIES}
\begin{figure*}[!htp]
	\centering
	{
		\includegraphics[width=1.0\textwidth]{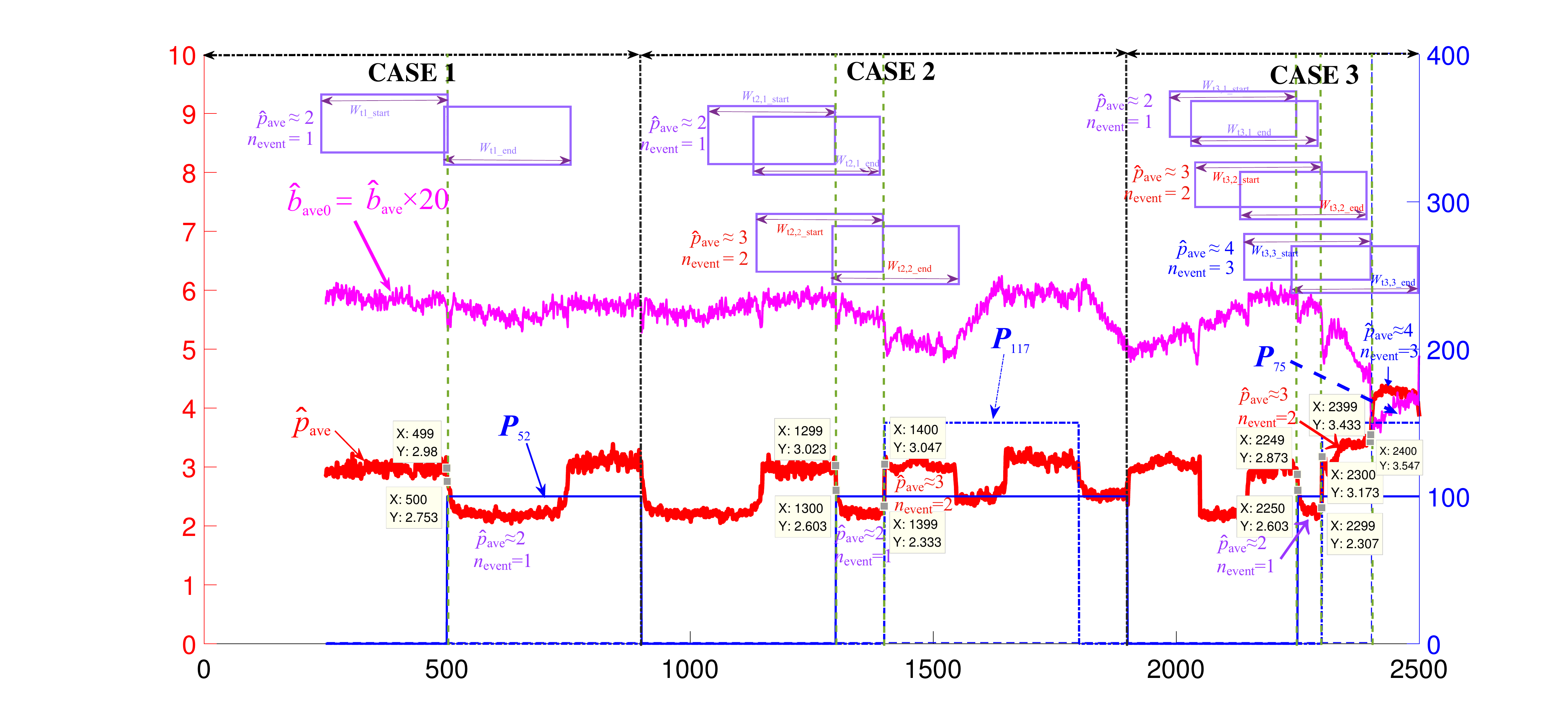}
	}
	\caption{{\label{fig2}} Case Stusies:  (a) Case 1 (b) Case 2 (c) in Case 3.}
\end{figure*}
The proposed method is tested with simulated data in the standard IEEE 118-bus system (the topology is shown in Fig. \ref{fig1} on the Matpower platform \cite{Zimmerman2011MATPOWER}. In this simulations, we regard a sudden power consumption (active power, $P$) change on some node as an event.

Three cases are designed to validate the proposed method. In case 1, case 2, case 3, we set different numbers of events on node 52, 117, 75, and observe the number of factors, respectively. To make a comparison, we illustrate the results of the three cases in the same picture as Fig. \ref{fig2}.

The raw data source, $\Omega$, is in size of $n$=118, $t$=2500. The size of the moving split-window  is set to be $N$=118, $T$=250, i.e. $\hat {\bm{X}} \in \mathbb{C} ^ {118\times250} $.

Then, \eqref{eq05} is used to estimate the parameters $p$ and $b$ as $\hat p$ and $\hat b$. The $p$ for the number of the assumed events, and the $b$ for the correlation structure of the noises. It is noted that we implement the simulated system model  for dozens of times to collect data, the noise of each time follows the same distribution. The dozens times simulation is reasonable, as that we can obtain dozens observations for a real physical system through its sampling data, which have noises following a certain distribtuion.

Then, in the $k$-th simulation, $\Omega_k$ is generated. With \eqref{eq05}, the estimation result  $\hat p_k$ and $\hat b_k$ are obtained. For these $\hat p_k$ and $\hat b_k$ $(k=1,2,\cdots)$, their mean value $\hat p_{\text{ave}}$ and $\hat b_{\text{ave}}$ is calculated, which may appear in decimal form. We need to point out that the estimation of $\hat p_{\text{ave}}$ and $\hat b_{\text{ave}}$ begins at ${t_s}$=250 due to the length of the split-window. In Fig.\ref{fig2}, we amplify the value of $\hat b_{\text{ave}}$ for twenty times to make it obvious, i.e. $\hat b_{\text{ave0}}=\hat b_{\text{ave}}\times20$.

The events in Case 1, Case 2, and Case 3 are given as Tab.\ref{table 3}, Tab.\ref{table 4}, and Tab.\ref{table 5}, respectively. The corresponding estimation resuts, i.e., the number of factors (i.e., $\hat p_{\text{ave}}$) and the correlation structure of the residuals (i.e., $\hat b_{\text{ave0}}$), are obtained as Fig. \ref{fig2}.

\subsection{Case 1: Single Event Detection}
\textit{Single Step Signal on Node 52:}

\begin{table}[ht]
	\begin{center}
		\caption{{\label{table 3}}  Events Assumed in Case 1}
		\begin{tabular}{p{0.8cm}|p{2.2cm}|p{3.5cm}}
			\toprule
			Node & Sampling Time & Power Consumption $P$ (MW)  \\
			\midrule
			52   &  ${t_s}=1\sim499$ & 0+$r_{52,t}$\\
			&  ${t_s} = 500 \sim 899$  & 100+$r_{52,t}$\\
			\midrule
			others   &  ${t_s} = 1 \sim 899$  & $c_k+r_{k,t}$\\
			\bottomrule
		\end{tabular}
		\raggedright
		
		{\small{}
			*$c_k$ is a constant of node $k$, $k=\{{1,2,\!\cdots\!,118}\}\!-\!\{52\}.$\\ *$r_{52,t}$,  $r_{k,t}$ are noises following AR(1) model, where $b_{\text{noise}}=0.5.$
		}
		\normalsize{}
	\end{center}
\end{table}

Fig. \ref{fig2}(a) shows that:
\begin{itemize}
	\item During the sampling time ${t_s}=250\!\sim\!499$\footnote{$250=1$(The beginning of Sigal)$+250$(Length of Split-Window)$-1$}. $\hat p_{\text{ave}}$ and $\hat b_{\text{ave}}$  remain steady around 3 and 0.28, respectively.
	\item At ${t_s}$=500, $\hat p_{\text{ave}}$ starts to decline to around 2. Also, $\hat b_{\text{ave}}$ declines slightly.
\end{itemize}

Actually, as can be seen from Tab. \ref{table 3}: no event occurs in the system when  $\hat p_{\text{ave}}$ keeps steady. Right at ${t_s}$=500,  the $P_{52}$ changes from 0 to 100MW.
Therefore, we can conduct  event detection with the proposed method.

Moreover, in this case, for the split-window $W_{t1}\!:\!t\in [251,500]$, there exist a single-event (i.e., step signal on Node 52, $\text{s}_{52}$, at ${t_s}$=500) and 2 factors, i.e., 1 event, $\hat p_{\text{ave}}\!\approx\!2$ during $W_{t1_\text{Start}}\!:\!t\in [251,500]$ to $W_{t1_\text{End}}\!:\!t\in [650,899]$. A linear relationship between the number of events and the number of factors will be revealed afterwards.

\subsection{Case 2: Multiple Event Detection (Two Events)}
\textit{Multiple Step Signal on Node 52 and Node 117:}

\begin{table}[ht]
	\begin{center}
		\caption{{\label{table 4}} Events Assumed in Case 2}
		\begin{tabular}{p{1cm}|p{3cm}|p{3cm}}
			\toprule
			Node & Sampling Time & Active Load (MW)  \\
			\midrule
			52   &  ${t_s} = 900 \sim 1299$ & 0+$r_{52,t}$ \\
			&  ${t_s} = 1300 \sim 1899$  & 100+$r_{52,t}$\\
			\midrule
			117   &  ${t_s} = 900 \sim 1399$ & 0+$r_{117,t}$ \\
			&  ${t_s} = 1400 \sim 1799$  & 150+$r_{117,t}$\\
			&  ${t_s} = 1800 \sim 1899$  & 0+$r_{117,t}$\\
			\midrule
			others   &  ${t_s} = 900 \sim 1899$   & $c_k+r_{k,t}$\\
			\bottomrule
		\end{tabular}
	\end{center}
\end{table}

Fig. \ref{fig2}(b) shows that:
\begin{itemize}
	\item During the sampling time ${t_s}=1149\!\sim\!1299$\footnote{$1149=900$(The beginning of Sigal)$+250$(Length of Split-Window)$-1$}, $\hat p_{\text{ave}}$ and $\hat b_{\text{ave}}$  remain steady. Thus, we deduce that no event occurs in the system, which meets Tab. \ref{table 4}.
	\item At ${t_s}$=1300, $\hat p_{\text{ave}}$ starts to decline (from 3.023 to 2.603) and then keeps around 2 till ${t_s}$=1399. For the split-window $W_{t2,1_\text{Start}}\!:\!t\in [1051,1300]$ to $W_{t2,1_\text{End}}\!:\!t\in [1150,1399]$, there exist a single-event (i.e., $\text{s}_{52}$ at ${t_s}$=1300) and 2 factors, i.e., 1 event, $\hat p_{\text{ave}}\!\approx\!2$.
	\item At ${t_s}$=1400, $\hat p_{\text{ave}}$ starts to raise (from 2.333 to 3.047) and then keeps around 3. For the split-window $W_{t2,2_\text{Start}}\!:\!t\in [1151,1400]$ to $W_{t2,2_\text{End}}\!:\!t\in [1299,1548]$, there exist two multi-event (i.e., $\text{s}_{52}$ at ${t_s}$=1300, $\text{s}_{117}$ at ${t_s}$=1400) and 3 factors, i.e., 2 event, $\hat p_{\text{ave}}\!\approx\!3$. 	
\end{itemize}


\subsection{Case 3: Multiple Event Detection (Three Events)}
\textit{Multiple Step Signal on Node 52, Node 117 and Node 75:}

\begin{table}[ht]
	\begin{center}
		\caption{{\label{table 5}} Events Assumed in Case 3}
		\begin{tabular}{p{1cm}|p{3cm}|p{3cm}}
			\toprule
			Node & Sampling Time & Active Load (MW)  \\
			\midrule
			52   &  ${t_s} = 1900 \sim 2249$ & 0+$r_{52,t}$ \\
			&  ${t_s} = 2250 \sim 2500$  & 100+$r_{52,t}$\\
			\midrule
			117   &  ${t_s} = 1900 \sim 2299$ & 0+$r_{117,t}$ \\
			&  ${t_s} = 2300 \sim 2500$  & 150+$r_{117,t}$\\
			\midrule
			75   &  ${t_s} = 1900 \sim 2399$ & 0+$r_{75,t}$ \\
			&  ${t_s} = 2400 \sim 2500$  & 400+$r_{75,t}$\\
			\midrule
			others   &  ${t_s} = 1900 \sim 2500$   & $c_k+r_{k,t}$\\
			\bottomrule
		\end{tabular}
	\end{center}
\end{table}

Fig. \ref{fig2}(c) shows that:
\begin{itemize}
	\item  During the sampling time ${t_s}=2149\!\sim\!2249$\footnote{$2149=1900$(The beginning of Sigal)$+250$(Length of Split-Window)$-1$}, $\hat p_{\text{ave}}$ and $\hat b_{\text{ave}}$  remain steady. Thus, we deduce that no event occurs in the system, which meets Tab. \ref{table 5}.
	\item At ${t_s}$=2250, $\hat p_{\text{ave}}$ starts to decline (from 2.873 to 2.603) and then keeps around 2 till ${t_s}$=2300. For the split-window $W_{t3,1_\text{Start}}\!:\!t\in [2001,2250]$ to $W_{t3,1_\text{End}}\!:\!t\in [2050,2299]$, there exist a single-event (i.e., $\text{s}_{52}$ at ${t_s}$=2250) and 2 factors, i.e., 1 event, $\hat p_{\text{ave}}\!\approx\!2$.
	\item At ${t_s}$=2300, $\hat p_{\text{ave}}$ starts to raise (from 2.307 to 3.173) and then keeps around 3 till ${t_s}$=2400. For the split-window $W_{t3,2_\text{Start}}\!:\!t\in [2051,2300]$ to $W_{t3,2_\text{End}}\!:\!t\in [2150,2399]$, there exist two multi-event (i.e., $\text{s}_{52}$ at ${t_s}$=2250, $\text{s}_{117}$ at ${t_s}$=2300) and 3 factors, i.e., 2 event, $\hat p_{\text{ave}}\!\approx\!3$.
	\item At ${t_s}$=2400, $\hat p_{\text{ave}}$ starts to raise (from 3.433 to 3.547) and then keeps around 4. For the split-window $W_{t3,3_\text{Start}}\!:\!t\in [2151,2400]$ to $W_{t3,3_\text{End}}\!:\!t\in [2251,2500]$, there exist three multi-event (i.e., $\text{s}_{52}$ at ${t_s}$=2250, $\text{s}_{117}$ at ${t_s}$=2300, $\text{s}_{75}$ at ${t_s}$=2400) and 4 factors, i.e., 3 event, $\hat p_{\text{ave}}\!\approx\!4$.
\end{itemize}
\subsection{Further Discussions about the Cases}
Through the above three cases, the relationship between the number of events (i.e., $\text{n}_{event}$) and the number of factors (i.e., $\hat p_{\text{ave}}$) is revealed.

The results of the three cases are summarized in Tab. \ref{table 6}:

\begin{table}[ht]
	\begin{center}
		\caption{{\label{table 6}} Relationship between Event Number ($\text{n}_{event}$) and Factor Number ($\hat p_{\text{ave}}$)}
		\begin{tabular}{p{0.5cm}|p{4cm}|p{1cm}|p{1cm}}
			\toprule
			Case & Split Window & ${n}_\text{event}$ & $\hat p_{\text{ave}}$ \\
			\midrule
			1  & $[251,500]\sim[650,899]$ & 1 & 2 \\
			\midrule
			2   &  $[1051,1300]\sim[1150,1399]$  & 1 & 2\\
			&  $[1151,1400]\sim[1299,1548]$  & 2 & 3\\
			\midrule
			3  	&  $[2001,2250]\sim[2050,2299]$  & 1 & 2\\
			&  $[2051,2300]\sim[2150,2399]$  & 2 & 3\\
			&  $[2151,2400]\sim[2251,2500]$  & 3 & 4\\
			\bottomrule
		\end{tabular}
	\end{center}
\end{table}

$\hat p_{\text{ave}}$, estimated by factor model, is approximately equal to ${n}_\text{event}$ plus one, i.e. $\hat p_{\text{ave}}\approx{n}_\text{event}+1$. There exists a linear relationship between them. Therefore, we can deduce the number of the multi-event for a certain split-window.

Besides, every time an event occurs,  $\hat b_{\text{ave}}$ drops. It indicates that the correlation in the residuals decreases when there exit events.

\section{CONCLUSION}
This paper proposes a data-driven method, namely, factor model, to conduct multi-event detection and recognition in a large power system. In the analysis procedure, we estimate the number of factors $p$ and the parameter for the correlation structure of residuals $b$ by minimizing the distance between two spectrums. Then, we conduct real-time analysis of the two parameters, $p$ and $b$, using moving split-window. The proposed method is direct and practical for multi-event analytics in a complex situation. Following conclusions are obtained: First, the number of factors estimated by factor model has an approximately linear relationship with the number of events that occur in the system. Second, taking non-Gaussian noises into account, time series  analytics is implemented to extract the information from noises. The decrease of parameter $b$ is related to the occurrence of events. It is noted that, the number of factors reveal the spatial information (events on different nodes) in the system; while the correlation structure in noises contains temporal information. The proposed method considers space-time correlation in a large power system. Finally, case studies verify the effectiveness of the method.

Along this direction, following work can be done. For example, we can employ more general modeling for noises. If we consider vector ARMA (1, 1) processes, we have up to 6th-order polynomial equations \cite{Zdzis2010A}. Furthermore, the relationship betweeen the number of factors and events can be further studied with physical model.

\appendices
\section{An Overview of Free Random Variable Techniques}
We summarize the main concepts and key results in free random variables techniques that we employ to derive ${\rho _{\bmod el}}(b)$. We follow the notations and derivations from \cite{Zdzis2010A,Burda2005Spectral}. First, consider a simple decomposition of covariance structures:
\begin{equation}
\label{eq12}
Co{v_{ia,jb}} = {A_{ij}}{B_{ab}}
\end{equation}

where $A$ is an $N \times N$ cross-covariance matrix and $B$ is a $T \times T$ auto-covariance matrix, $i,j = 1 \cdots N$, $a,b = 1 \cdots T$. Suppose $G$ is an $N \times T$ $i.i.d$ Gaussian random matrix. Then a correlated Gaussian random matrix $U$ ($N \times T$ time series) can be written as $U = {A_N}^{1/2}G{B_T}^{1/2}$. Its sample (empirical) covariance matrix $C$ is
\begin{equation}
\label{eq13}
C = \frac{1}{T}U{U^T} = \frac{1}{T}{A^{1/2}}GB{G^T}{A^{1/2}}
\end{equation}

Consider a real symmetric $N \times N$ random matrix $H$.\\
\textbf{\textit{Definition 1}} \textit{Mean Spectral Density}
\textit{
	\begin{equation}
	\label{eq14}
	{\rho _H}(\lambda ) = \frac{1}{N}\sum\limits_{i = 1}^N {\left\langle {\delta (\lambda  - {\lambda _i})} \right\rangle  = } \frac{1}{N}\left\langle {Tr(\lambda {1_N} - H)} \right\rangle
	\end{equation}}

where the expectation $\left\langle  \cdots  \right\rangle$ is taken w.r.t. the rotationally invariant probability measure, $\delta ( \bullet )$ is a Dirac delta function, and ${{1_N}}$ is a $N \times N$  unit matrix.\\
\textbf{\textit{Definition 2}} \textit{The Green’s Function (or Stieltjes Transform)}
\textit{
	\begin{center}
		${G_H}(z) = \frac{1}{N}\sum\limits_{i = 1}^N {\left\langle {\frac{1}{{z - {\lambda _i}}}} \right\rangle  = } \frac{1}{N}\left\langle {\frac{1}{{z{1_N} - H}}} \right\rangle $
	\end{center}
	\begin{equation}
	\label{eq15}
	 =\int {\frac{{{\rho _H}(\lambda )}}{{z - \lambda }}d\lambda }
	\end{equation}}

The relationship between ${\rho _H}(\lambda )$ and ${G_H}(z)$ is:
\begin{equation}
\label{eq16}
{\rho _H}(\lambda ) =  - \frac{1}{\pi }\mathop {\lim }\limits_{\varepsilon  \to {0^ + }} {\mathop{\rm Im}\nolimits} {G_H}(\lambda  + i\varepsilon )
\end{equation}

The Green’s function generates moments of a probability distribution, where the $n-th$ moment is defined by:\\
\textbf{\textit{Definition 3}} \textit{Moment}
\textit{
	\begin{equation}
	\label{eq17}
	{m_n} = \frac{1}{N}\left\langle {Tr{H^n}} \right\rangle  = \int {{\rho _H}(\lambda ){\lambda ^n}d\lambda }
	\end{equation}}
\textbf{\textit{Definition 4}} \textit{Moment Generating Function}
\textit{
	\begin{center}
		${G_H}(z) = \sum\limits_{n \ge 0} {\frac{{{m_n}}}{{{z^{n + 1}}}}}$
	\end{center}
	\begin{equation}
	\label{eq18}
	{M_H}(z) = \sum\limits_{n \ge 1} {\frac{{{m_n}}}{{{z^{n + 1}}}}}
	\end{equation}}

The relationship between ${G_H}(z)$ and ${M_H}(z)$ is
\begin{equation}
\label{eq23}
{M_H}(z) = z{G_H}(z) - 1
\end{equation}

Blue’s function and N-transform are the inverse transform of the Green’s function and moment generating function, respectively.\\
\textbf{\textit{Definition 5}} \textit{Blue’s function and N-transform}
\textit{
	\begin{center}
		${G_H}({B_H}(z)) = {B_H}({G_H}(z)) = z$
	\end{center}
	\begin{equation}
	\label{eq19}
    {M_H}({N_H}(z)) = {N_H}({M_H}(z)) = z
	\end{equation}}

Then, return to $Eq.13$. The N-transform of $C$ can be derived as:
\begin{equation}
\label{eq20}
{N_C}(z) = rz{N_B}(rz){N_A}(z)
\end{equation}

Using the moments’ generating function $M \equiv {M_C}(z)$ and its inverse relation to N-transform, $Eq.20$ can be written as:
\begin{equation}
\label{eq21}
z = rM{N_B}(rM){N_A}(M)
\end{equation}

Now, we consider the simplified model with ${A_N} \approx {I_{N \times N}}$. In such case, ${U_{nt}}$ is a time-series (AR(1)) following the autoregressive model:
\begin{equation}
\label{eq22}
{U_{nt}} = b{U_{n,t - 1}} + {\xi _{nt}}
\end{equation}

where $\left| b \right| < 1$, ${\xi _{nt}} \sim N(0,1 - {b^2})$, $n = 1, \cdots ,N,t = 1, \cdots ,T$. We calculate the eigenvalue distribution ${\rho _C}(\lambda )$ of correlation matrix $C = \frac{1}{T}U{U^T}$ based on the following strategy.\\
\textit{Step 1}: Find ${M_C}(z)$, from the equation for N-transform.\\
\textit{Step 2}: Find ${G_C}(z)$, by $Eq.19$.\\
\textit{Step 3}: Find ${\rho _C}(\lambda )$, by $Eq.16$.

For \textit{Step 1}, consider $Eq.22$. Because ${A_N} \approx {I_{N \times N}}$, so ${N_A}(z) = 1 + 1/z$. Therefore, $Eq.22$ can be rewritten as:
\begin{center}
	$\frac{z}{{r(1 + M)}} = {N_B}(rM)$
\end{center}
\begin{equation}
\label{eq24}
rM = {M_B}(\frac{z}{{r(1 + M)}})
\end{equation}

To find ${M_B}$, note that the auto-covariance matrix of AR(1) process has a simple form:
\begin{equation}
\label{eq25}
{B_{st}} = \frac{{{\mathop{\rm var}} (\zeta )}}{{1 - {b^2}}}{b^{\left| {s - t} \right|}} = {b^{\left| {s - t} \right|}}
\end{equation}

Using Fourier-transform of the matrix B, it can be shown that the moment generating function of B is
\begin{equation}
\label{eq26}
{M_B}(z) =  - \frac{1}{{\sqrt {1 - z} \sqrt {1 - \frac{{{{(1 + {b^2})}^2}}}{{1 - {b^2}}}z} }}
\end{equation}

Therefore, we obtain $Eq.10$ for \textit{Step 1}. The other steps are followed straightforwardly as $Eq.9$ and $Eq.8$.

\section{Kullback-Leibler divergence}
The Kullback-Leibler divergence is defined as follows:
\begin{equation}
\label{eq27}
{D_{KL}}(P\left\| Q \right.) = \sum\limits_i {{P_i}\log \frac{{{P_i}}}{{{Q_i}}}}
\end{equation}

where $P$ and $Q$ are probability densities. To deal with zero elements of $P$, we use:
\begin{equation}
\label{eq28}
{{\tilde P}_i} = \left\{ {\begin{array}{*{20}{c}}
	{\alpha {P_i}}\\
	\varepsilon
	\end{array}} \right.\begin{array}{*{20}{c}}
{,{P_i} > 0}\\
{,{P_i} = 0}
\end{array}
\end{equation}

where $\varepsilon$ is a small enough positive number. Denote the number of zero elements in $P$ as $num$, $\alpha  = 1 - num \times \varepsilon$. Probability density $Q$ is dealt with in the same way.

\bibliographystyle{IEEEtran}
\bibliography{YF}

\begin{thebibliography}{10}
\providecommand{\url}[1]{#1}
\csname url@samestyle\endcsname
\providecommand{\newblock}{\relax}
\providecommand{\bibinfo}[2]{#2}
\providecommand{\BIBentrySTDinterwordspacing}{\spaceskip=0pt\relax}
\providecommand{\BIBentryALTinterwordstretchfactor}{4}
\providecommand{\BIBentryALTinterwordspacing}{\spaceskip=\fontdimen2\font plus
\BIBentryALTinterwordstretchfactor\fontdimen3\font minus
  \fontdimen4\font\relax}
\providecommand{\BIBforeignlanguage}[2]{{%
\expandafter\ifx\csname l@#1\endcsname\relax
\typeout{** WARNING: IEEEtran.bst: No hyphenation pattern has been}%
\typeout{** loaded for the language `#1'. Using the pattern for}%
\typeout{** the default language instead.}%
\else
\language=\csname l@#1\endcsname
\fi
#2}}
\providecommand{\BIBdecl}{\relax}
\BIBdecl

\bibitem{Xu2015A}
X.~Xu, X.~He, Q.~Ai, and R.~C. Qiu, ``A correlation analysis method for power
  systems based on random matrix theory,'' \emph{IEEE Transactions on Smart
  Grid}, vol.~PP, no.~99, pp. 1--10, 2015.

\bibitem{Miao2014The}
X.~Miao and D.~Zhang, ``The opportunity and challenge of big data's application
  in distribution grids,'' in \emph{China International Conference on
  Electricity Distribution}, 2014, pp. 962--964.

\bibitem{Zhang20135Ws}
J.~Zhang and M.~L. Huang, ``5ws model for big data analysis and
  visualization,'' in \emph{IEEE International Conference on Computational
  Science and Engineering}, 2013, pp. 1021--1028.

\bibitem{Mayilvaganan2013A}
M.~Mayilvaganan and M.~Sabitha, \emph{A cloud-based architecture for Big-Data
  analytics in smart grid: A proposal}, 2013.

\bibitem{He2017A}
X.~He, Q.~Ai, R.~C. Qiu, W.~Huang, L.~Piao, and H.~Liu, ``A big data
  architecture design for smart grids based on random matrix theory,''
  \emph{IEEE Transactions on Smart Grid}, vol.~8, no.~2, pp. 674--686, 2017.

\bibitem{He2015An}
X.~He, R.~C. Qiu, Q.~Ai, and X.~Xu, ``An unsupervised learning method for early
  event detection in smart grid with big data,'' \emph{Computer Science}, 2015.

\bibitem{Chu2016Massive}
L.~Chu, R.~C. Qiu, X.~He, Z.~Ling, and Y.~Liu, ``Massive streaming pmu data
  modeling and analytics in smart grid state evaluation based on multiple
  high-dimensional covariance tests,'' \emph{IEEE Transactions on Big Data},
  vol.~PP, no.~99, pp. 1--1, 2016.

\bibitem{Smith2009Event}
M.~J. Smith and K.~Wedeward, ``Event detection and location in electric power
  systems using constrained optimization,'' in \emph{IEEE Power and Energy
  Society General Meeting}, 2009, pp. 1--6.

\bibitem{Soltan2014Cascading}
S.~Soltan, D.~Mazauric, and G.~Zussman, ``Cascading failures in power
  grids:analysis and algorithms,'' in \emph{International Conference on Future
  Energy Systems}, 2014, pp. 195--206.

\bibitem{He2011A}
M.~He and J.~Zhang, ``A dependency graph approach for fault detection and
  localization towards secure smart grid,'' \emph{IEEE Transactions on Smart
  Grid}, vol.~2, no.~2, pp. 342--351, 2011.

\bibitem{Xie2014Dimensionality}
L.~Xie, Y.~Chen, and P.~R. Kumar, ``Dimensionality reduction of synchrophasor
  data for early event detection: Linearized analysis,'' \emph{IEEE
  Transactions on Power Systems}, vol.~29, no.~6, pp. 2784--2794, 2014.

\bibitem{Rafferty2016Real}
M.~Rafferty, X.~Liu, D.~M. Laverty, and S.~Mcloone, ``Real-time multiple event
  detection and classification using moving window pca,'' \emph{IEEE
  Transactions on Smart Grid}, vol.~7, no.~5, pp. 2537--2548, 2016.

\bibitem{Wang2016Deep}
Y.~Wang, M.~Liu, and Z.~Bao, ``Deep learning neural network for power system
  fault diagnosis,'' in \emph{Control Conference}, 2016, pp. 6678--6683.

\bibitem{Naderian2016An}
S.~Naderian and A.~Salemnia, ``An implementation of type‐2 fuzzy kernel based
  support vector machine algorithm for power quality events classification,''
  \emph{International Transactions on Electrical Energy Systems}, vol.~27,
  no.~5, pp.~--, 2016.

\bibitem{Forni2016Dynamic}
M.~Forni, A.~Giovannelli, M.~Lippi, and S.~Soccorsi, ``Dynamic factor model
  with infinite dimensional factor space: Forecasting,'' \emph{Center for
  Economic Research}, 2016.

\bibitem{Yeo2016Random}
J.~Yeo and G.~Papanicolaou, ``Random matrix approach to estimation of
  high-dimensional factor models,'' \emph{Papers}, 2016.

\bibitem{Sun2016A}
Z.~Sun, X.~Liu, and L.~Wang, ``A hybrid segmentation method for multivariate
  time series based on the dynamic factor model,'' \emph{Stochastic
  Environmental Research and Risk Assessment}, pp. 1--14, 2016.

\bibitem{Wigner1951On}
E.~P. Wigner, ``On a class of analytic functions from the quantum theory of
  collisions,'' \emph{Annals of Mathematics}, vol.~53, no.~1, pp. 36--67, 1951.

\bibitem{Plerou2012Universal}
V.~Plerou, P.~Gopikrishnan, B.~Rosenow, L.~A.~N. Amaral, and H.~E. Stanley,
  ``Universal and non-universal properties of cross-correlations in financial
  time series,'' \emph{Papers}, vol.~83, no.~7, pp. 1471--1474, 2012.

\bibitem{LAURENT2000RANDOM}
L.~LALOUX, P.~CIZEAU, M.~POTTERS, and J.-P. BOUCHAUD, ``Random matrix theory
  and financial correlations,'' \emph{International Journal of Theoretical and
  Applied Finance}, vol.~3, no.~03, pp. 391--397, 2000.

\bibitem{Chevalier2016Identifying}
S.~C. Chevalier and P.~D.~H. Hines, ``Identifying system-wide early warning
  signs of instability in stochastic power systems,'' in \emph{Power and Energy
  Society General Meeting}, 2016, pp. 1--5.

\bibitem{Xu2005The}
C.~Xu, J.~Liang, Z.~Yun, and L.~Zhang, ``The small-disturbance voltage
  stability analysis through adaptive ar model based on pmu,'' in
  \emph{Transmission and Distribution Conference and Exhibition: Asia and
  Pacific, 2005 IEEE/PES}, 2005, pp. 1--5.

\bibitem{Zhang2007Spectral}
L.~Zhang, ``Spectral analysis of large dimentional random matrices,'' \emph{Ph
  D}, 2007.

\bibitem{Qiu2013Cognitive}
R.~Qiu and M.~Wicks, \emph{Cognitive Networked Sensing and Big Data}.\hskip 1em
  plus 0.5em minus 0.4em\relax Springer Publishing Company, Incorporated, 2013.

\bibitem{Donoho2000High}
D.~L. Donoho, ``High-dimensional data analysis: The curses and blessings of
  dimensionality,'' 2000.

\bibitem{Zdzis2010A}
Z.~Burda, A.~Jarosz, M.~A. Nowak, and M.~Snarska, ``A random matrix approach to
  varma processes,'' vol.~12, no. 1002.0934, pp. 1653--1655, 2010.

\bibitem{Zimmerman2011MATPOWER}
R.~D. Zimmerman and C.~E. Murillo-Sánchez, ``Matpower 4.1 user's manual,''
  \emph{Power Systems Engineering Research Center}, 2011.

\bibitem{Burda2005Spectral}
B.~Z, J.~J, and W.~B, ``Spectral moments of correlated wishart matrices.''
  \emph{Phys.rev.e}, vol.~71, no. 2 Pt 2, p. 026111, 2005.

\end{thebibliography}
\end{document}